\documentclass[submission, Phys]{SciPost}
\usepackage{amsmath}
\usepackage{amssymb}

\usepackage{epstopdf}
\usepackage{MnSymbol}
\usepackage{graphicx}
\usepackage{epsfig}
\usepackage{dcolumn}
\usepackage{bm}
\usepackage{color}
\usepackage{titlesec}
\usepackage{verbatim}

\def\(({\left(}
\def\)){\right)}
\def\[[{\left[}
\def\]]{\right]}

\newcommand{\beq}{\begin{equation}}
\newcommand{\eeq}{\end{equation}}
\newcommand{\barr}{\begin{eqnarray}}
\newcommand{\earr}{\end{eqnarray}}
\newcommand{\bei}{\begin{itemize}}
\newcommand{\eei}{\end{itemize}}

\newcommand{\e}{\text {e}}

\begin{document}

\begin{center}{\Large \textbf{
Annealed averages in spin and matrix models
}}\end{center}

\begin{center}
Laura~Foini\textsuperscript{1},
Jorge~Kurchan\textsuperscript{2}
\end{center}

\begin{center}
{\bf 1} IPhT, CNRS, CEA, Universit\'{e} Paris Saclay, 91191 Gif-sur-Yvette, France
\\
{\bf 2} Laboratoire de Physique de l'ENS, Ecole Normale Sup\'erieure, PSL Research University, Universit\'e Paris Diderot, Sorbonne Paris Cit\'e, Sorbonne Universit\'es, UPMC Univ. Paris 06, CNRS, 75005 Paris, France
\\
* laura.foini@ipht.fr
\end{center}

\begin{center}
\today
\end{center}

\section*{Abstract}
{\bf
A disordered system is  denominated `annealed' when the interactions 
themselves may evolve and adjust their
values to lower the free energy. 
The opposite  (`quenched') situation when disorder is fixed, is the
one relevant for physical spin-glasses, and has received vastly more attention. 
Other problems however are more natural in the annealed situation:
in this work we discuss examples where annealed averages 
are  interesting, in the context of 
matrix models.
 We first discuss how in practice, when  system
 and  disorder adapt together,
  annealed systems develop `planted' solutions spontaneously, as the ones found in the 
  study of inference problems.
 In the second part, we study  the probability distribution of elements of a matrix derived from a 
rotationally invariant (not necessarily Gaussian) ensemble,  a problem that maps into the annealed average of
a spin glass model.  }

\vspace{10pt}
\noindent\rule{\textwidth}{1pt}
\tableofcontents\thispagestyle{fancy}
\noindent\rule{\textwidth}{1pt}
\vspace{10pt}

\section{Introduction}

Consider the following problem: we are given a system of  size $N$ depending on disorder variables $J$, and a  set of distributions  $P_J$,  that  we expect to depend exponentially in $N$. The most usual example is a thermodynamic system of, say, $N$ spins with random interactions $J_{ij}$. The corresponding
distribution of energies of configurations  $e^{S_J(E)}\propto P_J(E) $ defines  the (disorder-dependent) entropy.
We now consider many samples, and wish to average them: we may choose a {\em quenched} average $\langle \ln P_J(E) \rangle_J$ or an 
{\em annealed} average $\ln \langle P_J(E) \rangle_J$. The two results are generally different \cite{mezard1987spin,cavagna_2005}.
In physical, finite-dimensional systems  with short-range interactions,  one can
imagine the system as being composed of many quasi-independent parts, and conclude that the free-energy, energy and entropy (but not 
their exponentials) are the addition of
their values in these parts.
This argument  suggests that these are the quantities that have to be averaged, since they are the ones
that concentrate in probability in the observed value, their average that gives the typical results for one sample.

 Annealed averages apply when the disorder evolves in equilibrium with the spins, so there is no reason that they
 should be treated differently. They have  been less studied, but there are reasons to try to understand them better. 
The origin of our motivation is the following problem: we are given a large $N\times N$ matrix $A$, drawn from a rotationally invariant
(generically non-Gaussian) distribution ${\cal{P}}(A) = {\cal{P}}(U A U^\dag)$, where $U$ may be orthogonal, unitary or symplectic \cite{mehta2004random,livan2018introduction}. 
 A well-studied example is a `matrix  model' ${\cal{P}}(A) \sim e^{- N {\mbox{tr}} W(A)}$, for some potential $W$ \cite{brezin1993planar}.
We wish, for example, to know what is the probability marginal  distribution of the diagonal $P_{A_{ii}}$
(or equivalently $P_{{\boldsymbol \sigma} \cdot A {\boldsymbol \sigma}}$ with ${\boldsymbol \sigma}$ a $N$-component vector), or of the off-diagonal  element $P_{A_{ij}}$. 
Now, as it is well-known, the integration over all random matrices $A$ may be split in an integration over eigenvalues and another over `angle'  variables defining the eigenbasis.
As we shall see, making a quenched calculation  amounts to  treating  eigenvalues of the disorder matrix as fixed and eigenvectors as annealed, although they are  originally variables on an equal footing describing the matrix $A$. 
In conclusion, this is an instance in which the annealed calculation seems the natural one. As we shall see, the result of both approaches differ,
and this shows up in the large deviations of the probability distributions.

Our study of the annealed average of spin-glass  models shows that the freedom  of the couplings
to adapt to the spin configurations leads at low temperatures to self-planted solutions~\cite{zdeborova2016statistical}.
These are configurations with particular low energy with respect to the quenched ones as a result of the annealing.
The understanding of this phenomenon that we review and discuss in detail in the first part of our work,
is somehow scattered in the literature. In more detail,
Hidden Mattis phases, a particular instance of this,  have indeed been discussed long ago (see section \ref{Sec_stat_phys})  in \cite{kasai1983hidden,matsuda1984hidden}
and the idea of self-planting due to time evolving disorder has been pointed out more recently in the perceptron model \cite{sharma2019self}.
Slowly varying interactions with applications to physics and biology have also been considered in \cite{penney1993coupled,rabello2008solvable}.
There is also a closely related computation of Dean and Majumdar \cite{dean2006large,dean2008extreme}, which involves the
large deviations of the lowest eigenvalue of a Gaussian matrix: here we are interested on those towards lower values, the one towards 
higher values is not relevant here. Yet another example that is easy to understand is the high-pressure phase  of spheres with polydispersity. If polydispersity is left to vary freely, each particle will expand as much as allowed, giving rise to a packing with very different statistical properties
(see \cite{brito2018universality,ikeda2020jamming}).

In the first part of this work (section \ref{Sec_stat_phys}) we discuss in general 
the global picture which emerges when one considers annealed averages
in statistical models with disorder. 
In the second part (section \ref{Sec_matrix})  we proceed to calculate the (annealed) joint distribution of an $r \times r$ submatrix ($r$  finite) of a large $N\times N$ random matrix
derived from a general rotationally invariant  matrix model.

\section{Annealed averages in statistical models}\label{Sec_stat_phys}

In this part of the paper we address the question of what happens to a model when we allow its disorder to adapt, i.e. when we perform
an annealed average. Some of the main features may be seen more clearly in the simple case of spherical models, which we shall review first. 

\subsection{Spherical SK model}\label{Sec_spherical}

Working with continuous spins, we have at our disposal the possibility of solving the problem
diagonalizing the interaction matrix, so that the discussion is particularly simple.
The partition function  reads:
\beq\label{Eq_p2}
Z_J(\beta) = \int {\rm d} \mathbf{s} \ e^{\frac12 \beta  \sum_{ij}J_{ij} s_i s_j } \delta\left(\sum_i s_i^2 - N\right)
= \int  {\rm d} \tilde{\mathbf{s}} \ e^{\frac12 \beta  \sum_{k} \lambda_k \tilde{s}_k^2} \delta\left(\sum_i \tilde{s}_i^2 - N\right) \ ,
\eeq
with $ {\rm d} \mathbf{s} = \prod_{i}^N {\rm d} s_i$ and 
where in the second step we have diagonalized the Hamiltonian.
We consider here the case of $J$ real and symmetric.
For a rotationally invariant orthogonal ensemble with potential ${\cal P}(J) \sim e^{- \frac{N}{2} \text{tr} V(J)}$ 
the eigenvalues $\lambda_i$ are distributed according to the energy $N^2 E({\boldsymbol \lambda})$ with:
\begin{equation}
E({\boldsymbol \lambda}) =  \frac{1}{2 N} \sum_i V(\lambda_i) -  \frac{1}{N^2} \sum_{i>j} \log|\lambda_i-\lambda_j|
\end{equation}

Note that both terms in the action $E({\boldsymbol \lambda})$ are of order one and one can hope for a non trivial 
large $N$ limit.
As mentioned in the Introduction, there are two possibilities to perform the average over the coupling matrix $J$ (or its eigenvalues).
From the statistical physics point of view, one is usually interested in the quenched average of the partition function. 
Here we do both in preparation for the second part of the paper dedicated to  matrix models, where annealed averages are more relevant
(for  discussion of quenched versus annealed averages in random matrix ensembles see also
Ref \cite{potters2019first}).

\begin{figure}
\centering \includegraphics[angle=0,width=7cm]{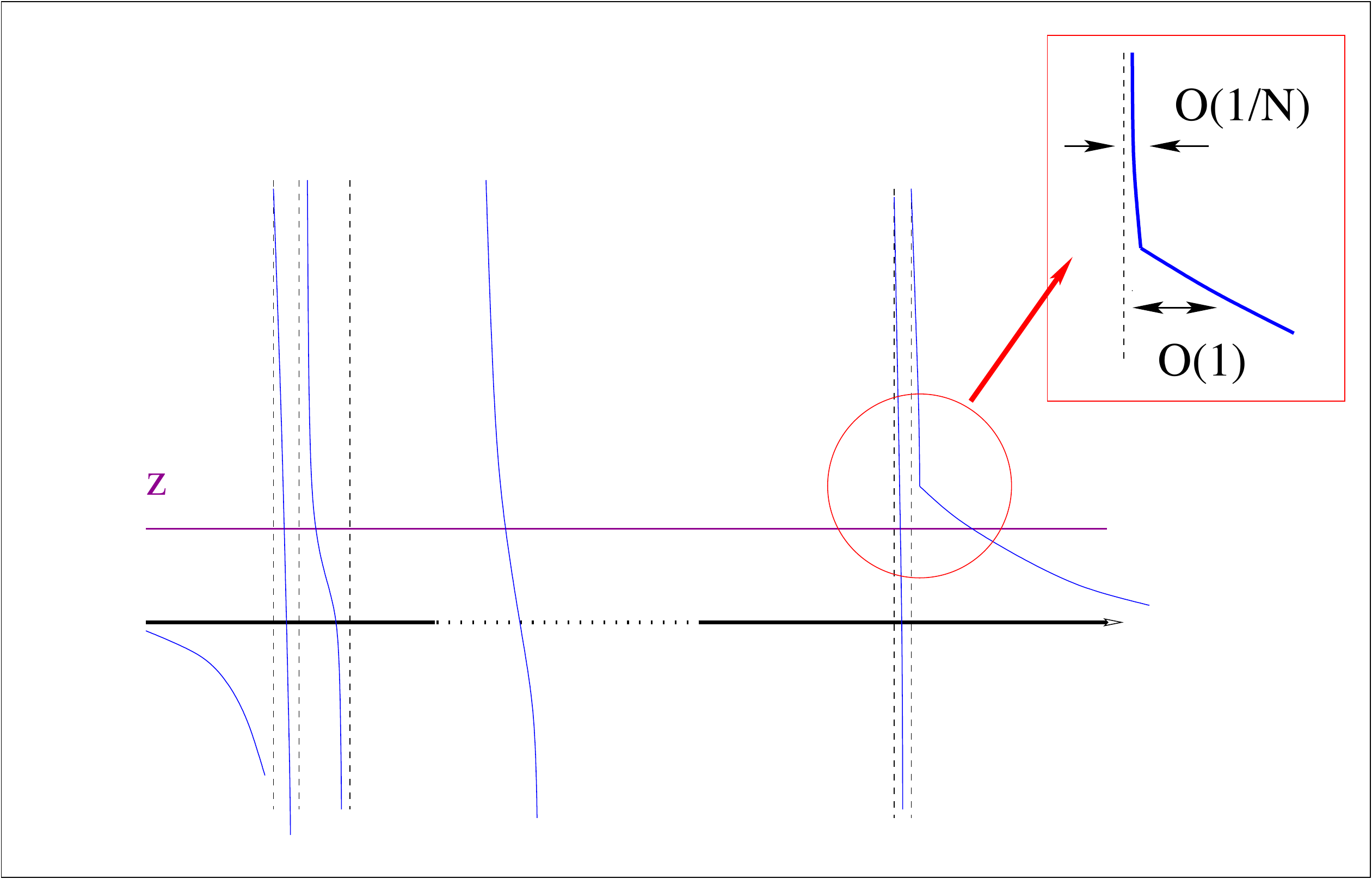}
\caption{The dispersion relation and the mechanism of freezing: in the large $N$ limit the curve has, to the right of the last pole,  a jump in the derivative.}
\label{dispersion}
 \end{figure}

\begin{itemize}
\item {\em Quenched average.}
We first draw the $\lambda_i$ from the probability distribution $P({\boldsymbol \lambda}) \propto  e^{-E({\boldsymbol \lambda})}$.
For example, in a Gaussian ensemble this leads to the semicircle distribution for the $\lambda_i$, for large $N$ \cite{mehta2004random}. 
Then, at fixed $\lambda$, we compute expectations of functions of the $s_i$ based on (\ref{Eq_p2})  \cite{kosterlitz1976}. 
In practice, this procedure leads to the following average
\beq\label{Eq_p2_0}
\begin{array}{ll}
\displaystyle  \langle \ln Z_J(\beta)  \rangle & 
\displaystyle = \int  {\rm d} \boldsymbol{ \lambda} e^{-N^2 E({\boldsymbol \lambda}) }\ln \left\{  \int  {\rm D} \tilde{\mathbf{s}}  \; \e^{ \frac12 \beta  \sum_{k} \lambda_k \tilde{s}_k \tilde{s}_k } \;  \delta\left(\sum_i \tilde{s}_k^2 - N\right) \right\} \ .
\end{array}
\eeq
Note that the average over the spins $\tilde s_i$, and consequently on the eigenvectors, is annealed.
The logic is simple: the distribution of the spins, however peaked, will not distort the measure of the eigenvalues in the large $N$-limit, because it is under a logarithm.

\item {\em Annealed  average.}  We treat the $\lambda_i$ and the $\tilde s_i$ on an equal footing, so that we consider
the combined measure:
\beq\label{Eq_p2_2}
\begin{array}{ll}
\displaystyle  \langle Z_J(\beta)  \rangle & 
\displaystyle = \int  {\rm d} \boldsymbol{ \lambda}  {\rm d} \tilde{\mathbf{s}} \ {\rm d} z \; \e^{-N^2 E({\boldsymbol \lambda}) + \frac12 \beta  \sum_{k} \lambda_k \tilde{s}_k \tilde{s}_k } \;  \delta\left(\sum_i \tilde{s}_k^2 - N\right) \ .
\end{array}
\eeq

\subsection{The electrostatic analogy}\label{Sec_electro}

Introducing a Lagrange multiplier in (\ref{Eq_p2}) we obtain:
\beq\label{Eq_p2_2a}
\begin{array}{ll}
\displaystyle  Z_J(\beta)  & 
\displaystyle = \int {\rm d} \tilde{\mathbf{s}} \ {\rm d} z \ e^{\frac12 \beta  \sum_{k} \lambda_k \tilde{s}_k \tilde{s}_k - \frac{z}{2}  \sum_k \tilde{s}_k^2 } e^{\frac{N z}{2}} 
 \vspace{-0.2cm} 
 = \displaystyle \int {\rm d} z \ e^{ \frac{N}{2} \left[z + \log 2\pi - \frac{1}{N}  \sum_{k} \log ( z - \beta \lambda_k)  \right] }  \ .
\end{array}
\eeq
In the following we will make the change of variable $z\to \beta z$.

For the {\em quenched } solution we have to solve first for the $\lambda_i$:
\begin{equation}\label{Eq_eigenvalues_quenched}
V'(\lambda_i) -  \frac{2}{N} \sum_{j (\neq i)} \frac{1}{\lambda_i-\lambda_j}=0 \ ,
\end{equation}
a classical exercise in random matrix theory. Then, at fixed $\lambda_i$ we need to solve the dispersion equation:
\beq\label{eq_z}
\beta = \frac{1}{N} \sum_k \frac{1}{z-\lambda_k} 
\eeq
to determine $z$.
We may think (\ref{eq_z})  as an electrostatic equation in one dimension. We need to find the value of $z$ 
where the `electric field' reaches the value $\beta = \frac 1 T$. At small values of $T$, the solution is a continuous branch. Depending on the distribution of eigenvalues, in particular for  a semicircle distribution, the
`electric field' reaches the vicinity of the rightmost pole with a finite value, and  grows sharply at distance $O(1/N)$ of the last pole. This situation is depicted in Figure \ref{dispersion}.

From the point of view of large-$N$, this is often described by saying that  the root `sticks' at the value of the largest pole. 
Note that this situation of finite field up to
very close to a system of charges, and then divergence close to the charges,  is the usual one in a charged metal!
If the `charge density' instead falls at the edge sufficiently fast, then the `electric field' does not reach a finite limit at distances $O(1)$ from the edge and there is no freezing mechanism, and hence no transition. An example of this is a constant density of eigenvalues.

\end{itemize}

\begin{figure}
\centering \includegraphics[angle=0,width=7cm]{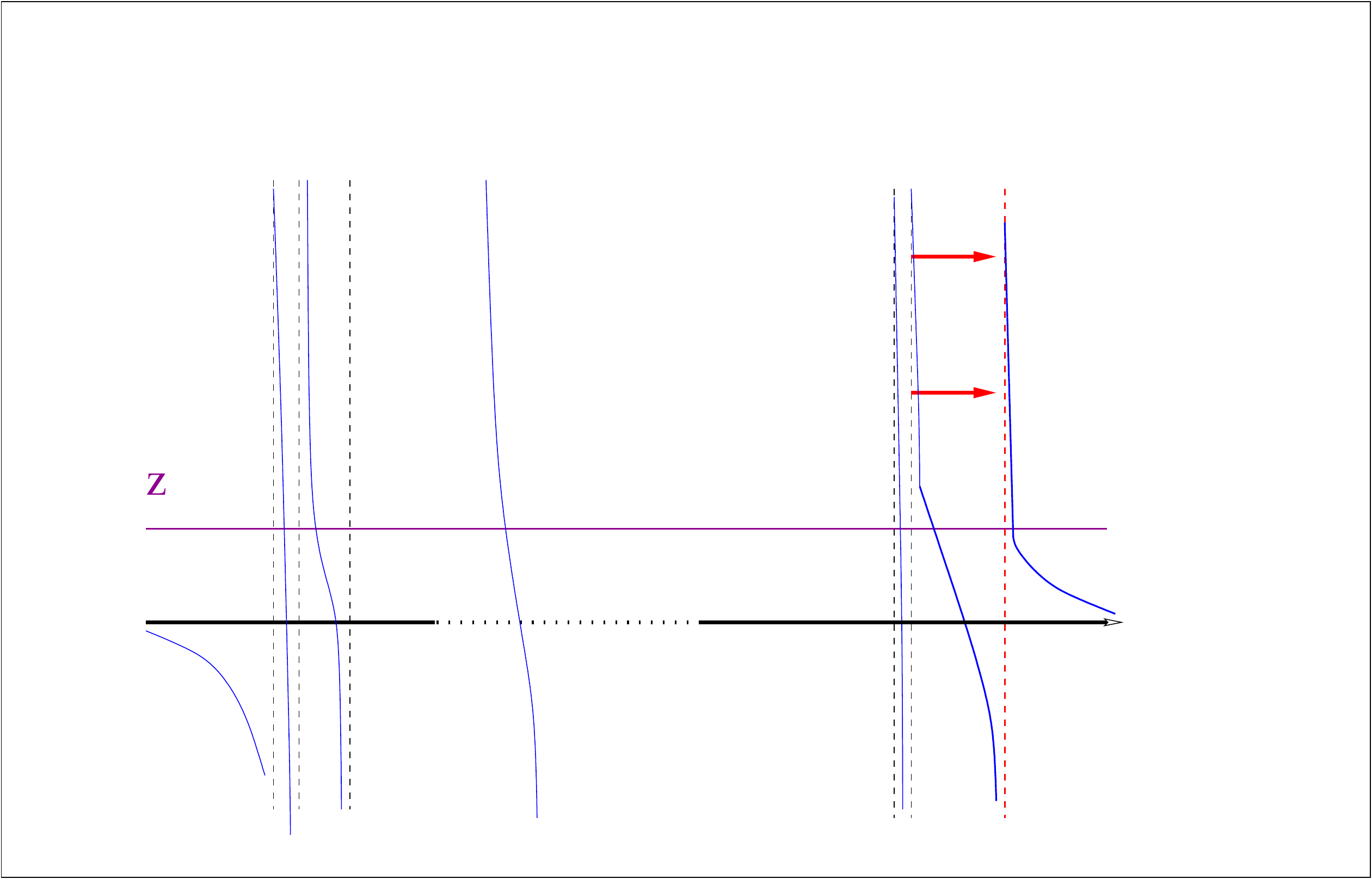}
\caption{Annealed case: the transition is avoided by detaching an eigenvalue, which stays close ($O(1/N)$) to the value of $z$ }
\label{dispersion1}
 \end{figure}

For the {\em annealed} solution we need to minimize:
\begin{equation}\label{Eq_minimization_annealed}
 \frac{1}{2} \sum_i V(\lambda_i) - \frac{1}{N} \sum_{i > j} \log|\lambda_i-\lambda_j| - \frac{\beta}{2} z  + \frac{1}{2N}  \sum_{k} \log ( z -  \lambda_k)  + {\mbox{const}}
\end{equation}
We have now $N$  negative charges  
and one positive charge  at $(\lambda_i,z)$ which have to be treated on an equal footing. 
Note however that $V'$ acts only on the $\lambda_i$ charges, while there
is a linear potential only on  $z$.
For large temperatures the solution is the same as in (\ref{Eq_eigenvalues_quenched}), because then the charge at $z$ is far and, being a single one,  has negligible effect on the bulk.
When $T$ is such that $z$ approaches the last charge of the bulk, something remarkable happens: positive and negative charges ($z,\lambda_N$) form a ``molecule" of ``size"  $|z-\lambda_N| = O(1/N)$ which breaks loose from the bulk and moves to the right.
The molecule is subject to interaction  with $V$ via $\lambda_N$, and with the linear potential via $ z$. The solution where these forces cancel  is depicted in Fig. \ref{dispersion1}. 

From the point of view of the interactions $J$ one has a detached eigenvalue and therefore the coupling matrix can be decomposed in $\sum J_{ij} s_i s_j= \sum J_{ij}' s_i s_j + \lambda_N (\sum_i v_i s_i)^2 $, where $J'$ is a matrix very similar to the unperturbed
one, and ${\bf v}$ is the normalized eigenvector with the detached eigenvalue $\lambda_N$. This last term constitutes a ferromagnetic (or rather `Mattis') term \cite{mattis1976solvable}. This is
the mechanism developed by the interaction to lower the free energy.
Note also that this form of the coupling matrix is the one used in the so-called planted ensemble for the 
study of inference problems, where the vector $\boldsymbol{v}$ represents the signal that one aims to recover \cite{zdeborova2016statistical}.
In this context in fact, it is known that a rank one perturbation can shift the largest eigenvalue of the original matrix \cite{baik2005phase}.

\subsection{The solution in terms of the R-transform}

Here we describe a  calculation that shows explicitly how the mechanism of detaching an eigenvalue occurs 
for the spherical model and that 
  the annealed free entropy (\ref{Eq_p2_2}) is analytic at all temperatures.
We will make use of the Stieltjes transform which is defined as
\beq
S(z) = \frac{1}{N} \sum_{k=1}^N \frac{1}{z-\lambda_k} = \int \frac{\rho(\lambda)}{z-\lambda} {\rm d} \lambda
\eeq
in the limit of large $N$ and
with $\rho(\lambda)$ the (averaged) asymptotic eigenvalue distribution of the matrix $J$.
The R-transform is defined from the inverse of the Stieltjes
\beq\label{Eq_R_Stiel}
R(\omega) = S^{-1}(\omega) - \frac{1}{\omega} \ .
\eeq
The R-transform admits a representation in terms of a series expansion where the coefficients
can be determined explicitly and are called free cumulants \cite{tulino2004random}:
\beq\label{Eq_R_series}
R(\omega) = \sum_{k=1}^{\infty} C_k \omega^{k-1}
\eeq
From (\ref{Eq_R_Stiel}) 
the R-transform is defined on the real axes in $\omega \in S([\lambda_-,\lambda_+]^c)$, 
with $\lambda_\pm$ the boundary of the support of the density of eigenvalues,
but one can consider its analytical continuation as a complex function \cite{kosterlitz1976,tulino2004random}.
Its radius
of convergence in (\ref{Eq_R_series}) can be larger and  can be continued beyond $\omega=S(\lambda_+)$.

Let us first see the result at high temperatures for the quenched calculation (\ref{Eq_p2_0}).
In terms of these transforms Eq. (\ref{eq_z}) 
reads $z = S^{-1}(\beta) =  R(\beta)+\beta^{-1}$. 
This solution is valid when $z \geq \lambda_+$ namely $S(\lambda_+) \geq \beta$.
For high temperature one can therefore conclude
that the quantity $\Phi = \frac{2}{N} \log Z$ is given by \cite{marinari1994replica,guionnet2005fourier,maillard2019high,BBP_2017}:
\beq\label{High_T}
\Phi \sim
\int_0^\beta R_{\rho}(x) {\rm d} x
\eeq
As we discussed above,  at $\beta_c$ there is a phase transition where the Lagrange multiplier $z$ sticks on the boundary of the spectrum \cite{kosterlitz1976}
and consequently the free energy has a non-analiticity.

In the  annealed computation
 one minimizes at the same time over $z$ and over the eigenvalues 
the function (\ref{Eq_minimization_annealed}).
Assuming as above  that at low temperatures the eigenvalue $\lambda_N$ separates from the bulk and $z = \lambda_N + \frac{1}{N}  a(\lambda_N)$,
 the eigenvalues in the bulk satisfy the following equations \cite{livan2018introduction,brezin1993planar}:
\beq
 V'(\lambda_i) = \ 2 \ p.v. \int {\rm d} \lambda \frac{\rho(\lambda)}{\lambda_i-\lambda} + O(1/N)
\eeq
for $i=1,\dots,N-1$, 
where $p.v.$ stands for principal value, 
while for the external eigenvalue $\lambda_N$:
\beq
 V'(\lambda_N) =   2 S(\lambda_N) + \frac{1}{a(\lambda_N)} \ .
\eeq
In order to make the mechanism of the annealed solution explicit, we restrict here to the case in which the potential is Gaussian, namely $V(x)=x^2/2$ 
because in that case it is possible to get explicit expressions for $S$ and $R$
but  we believe it applies to a much wider class
of rotationally invariant ensembles, as most of our derivation is quite general. 
For a matrix ensemble with a compact support the Stieltjes transform takes the form~\cite{brezin1993planar,BBP_2017}:
\beq\label{Eq_st}
S_{\pm}(x) =  \frac{1}{2} V'(x) \pm Q(x) \sqrt{(x-\lambda_+)(x-\lambda_-)} 
\eeq
where $Q(x)$ is a polynomial and $\lambda_{\pm}$ are the boundaries of the support of the
density of eigenvalues. 
The sign in front of the square root in (\ref{Eq_st}) is chosen such that $S(x) \to 1/x$ for $|x|\to\infty$.
Assuming that the minus sign describes the solution at large positive $x$,
this implies that
\beq
 \frac{1}{a(\lambda_N)} =  2 Q(\lambda_N) \sqrt{(\lambda_N-\lambda_+)(\lambda_N-\lambda_-)} \ .
\eeq
which also implies that one can read the solution on the ``non-physical" branch of the Stieltjes transform because:
\beq
S_-(\lambda_N) +  \frac{1}{a(\lambda_N)}  = S_+(\lambda_N) 
\eeq
where $S_-=S$ at the right of the support of the spectral density.
In fact this is the equation satisfied by the Lagrange multiplier $z = \lambda_N + \frac{1}{N} a(\lambda_N)$:
\beq
\beta = S_-(\lambda_N) + \frac{1}{a(\lambda_N)} = S_+(\lambda_N) 
\eeq
which implies that for the annealed calculation the solution (\ref{High_T}) holds for all temperatures
because $S_+^{-1}$ is the continuation of $S^{-1}$ at larger values of $\beta$.

\begin{figure}
\centering \includegraphics[angle=0,width=7cm]{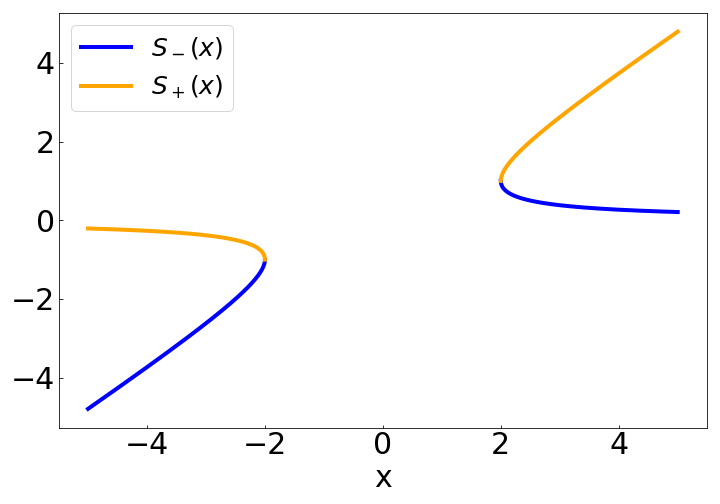}
\caption{Stieltjes transform for the Gaussian ensemble $S_{\pm}(x) = \frac12 (x\pm\sqrt{x^2-4})$.
The density of eigenvalues is defined between $[-2,2]$.
The physical solution is the one that goes to zero as $1/x$ in the limit $|x| \to \infty$. 
However its continuation on the other side of the support of the density of eigenvalues
intervenes in the solution of the annealed problem.}
\label{Stieltjes}
 \end{figure}

The point of this calculation is that the complex mechanism of detachment of a single eigenvalue and formation of a `molecule' 
reduces in this formalism to just following the `unphysical' branch of the Stieltjes transform, 
as shown in Figure \ref{Stieltjes}. Understanding this better in the most general non-Gaussian case deserves a deeper analysis.

\subsection{Two  energies and two (un)constrained replicas }\label{Two_replicas_two_T}

In this section we consider the annealed solution of the same problem,  with two sets of spins and two energies. This turns out
to be important for the discussion of matrix elements in random matrix models 
in the second part of the work.
At this point it is important to specify whether the two replicas are free to
overlap, or they are forced to be orthogonal. The latter  will be the case for the computation in matrix models Section \ref{Sec_diagonal}.
The presence of this constraint leads to two different solutions.

We consider the following annealed average:
\beq
\langle Z_J(\beta_1,\beta_2)\rangle = \int {\rm d} J \ P(J) \int {\rm d} \mathbf{s}^1  \; {\rm d} \mathbf{s}^2
e^{ \beta_1 \sum_{i<j} J_{ij} s_i^1 s_j^1+ \beta_2 \sum_{i<j} J_{ij} s_i^2 s_j^2}\;
\delta\left(\sum_i s^1_i - N \right) \delta\left(\sum_i s^2_i - N \right)
\eeq

In the case where the two replicas are constrained to be orthogonal at low enough
temperature two eigenvalues detach from the bulk. 
If this constraint is not imposed it leads to a different solution.
To see this we introduce Lagrange multipliers and we proceed in a similar way as before:
\beq
\begin{array}{ll}
\displaystyle
\langle Z_J(\beta_1,\beta_2)\rangle & \displaystyle = \int {\rm d} \boldsymbol{ \lambda} e^{-N^2 E({\boldsymbol \lambda}) } \int {\rm d} \tilde{\mathbf{s}}^1  \int {\rm d} \tilde{\mathbf{s}}^2
e^{ \frac12 \beta_1 \sum_{k} \lambda_k (\tilde{s}_k^1)^2  +  \frac12 \beta_2 \sum_{k} \lambda_k (\tilde{s}_k^2)^2 
-  \frac12 \beta_1 z_1 ( \sum_k \tilde{s}_k^1 - N) -  \frac12 \beta_2 z_2 ( \sum_k \tilde{s}_k^2 - N)}
\\ \vspace{-0.2cm} \\
& \propto \displaystyle
\int  {\rm d} \boldsymbol{ \lambda} e^{- N^2 E({\boldsymbol \lambda}) }
e^{\frac{N}{2} \left[ - \frac{1}{N} \sum_k \log(z_1-\lambda_k)  - \frac{1}{N} \sum_k \log(z_2-\lambda_k)
+ \beta_1 z_1 +\beta_2 z_2 \right] }
\end{array}
\eeq

Our electrostatic problem has now $N+2$ charges. 
We need to minimize:
\begin{equation}
 \frac{1}{2} \sum_i V(\lambda_i) -  \frac1{N}\sum_{i> j} \log|\lambda_i-\lambda_j| - \frac{1}{2} (\beta_1 z_1+ \beta_2 z_2)  + \frac{1}{2N}    \sum_{k} \log ( z_1 -  \lambda_k)  + \frac{1}{2N}  \sum_{k} \log ( z_2 -  \lambda_k) \ .
\end{equation}

The saddle point over $z_1$ and $z_2$ gives:
\beq
\beta_i = \frac{1}{N} \sum_k \frac{1}{z_i-\lambda_k} \ .
\eeq

The solution  at high temperatures $\beta_1$ and $\beta_2$ is as two independent replicas
with the unperturbed set of eigenvalues for $J$.
For low enough temperatures, {\em both positive},   one eigenvalue detaches for the bulk and both $z_i$
multipliers attach to it,  forming a   three-component molecule $(\lambda_N,z_1,z_2)$ of size $1/N$.
For temperatures close to zero {\em but of opposite sign}, there is one eigenvalue at each extreme of the bulk that detaches: the system
has a ``molecule" to the right $(z_1,\lambda_N)$, and another $(z_2,\lambda_1)$ to the left of the bulk.
The situation with opposite temperatures is relevant for the computation
of off-diagonal matrix elements in matrix models (see Section \ref{Sec_off}).

\subsection{The solution of two unconstrained replicas in terms of the R-transform}

Let us see in detail the case of positive, low temperatures.
At high temperature the eigenvalue distribution is not modified and the result of the partition function is
that of two independent replicas:
\beq\label{Fo}
\langle Z_J(\beta_1,\beta_2)\rangle \propto e^{\frac{N}{2} \left( \int_0^{\beta_1} R(x) {\rm d} x +  \int_0^{\beta_2} R(x) {\rm d} x \right)} \ .
\eeq
We first assume now that both $\beta_1$ and $\beta_2$ are large and both 
Lagrange multipliers are close to the detached eigenvalue $\lambda_N$, namely
 $z_1 =  \lambda_N + a_1/N$ and $z_2 =  \lambda_N + a_2/N$.
This ansatz translates into:
\beq\label{Eq_lagr}
\beta_i = S(\lambda_N) + \frac{1}{a_i} \ .
\eeq
and the equation for the largest eigenvalue reads:
\beq
V'(\lambda_N) = 2 S(\lambda_N) + \frac{1}{a_1} + \frac{1}{a_2}
\eeq
which combined with (\ref{Eq_lagr}) gives:
\beq
\beta_1+\beta_2 = V'(\lambda_N) \ .
\eeq
Assuming now that one eigenvalue $\lambda_N$ has detached and that only one 
Lagrange multiplier is close to it, which occurs when
$\beta_1 > S(\lambda_{N})$ and $\beta_2 < S(\lambda_{N})$ (or the same with exchange of $\beta_1$ and $\beta_2$),
the second Lagrange multipliers sticks to it under the condition:
\beq
\beta_1 = S_+(\lambda_N) \qquad \beta_2 = S(\lambda_N) \ .
\eeq
In particular for the Gaussian ensemble this implies $\beta_1\beta_2 = 1$.
Note that at this transition point the condition of analyticity is not ensured,
as it was also found in \cite{gardner1989}.   

Let us now see the case of opposite temperatures $\beta_1>0$ and $\beta_2<0$.
At high temperature the partition function is again (\ref{Fo}) and the spectrum
of the eigenvalues is unperturbed.
However one has to be careful with the continuation of this result to small temperatures.
In this limit two eigenvalues detaches from the boundary left and right
when
\beq
\beta_1 = S(\lambda_N) \qquad \text{and}\qquad \beta_2 = S(\lambda_1)
\eeq
We suppose as before that $z_1 =  \lambda_N + \frac{1}{N}  a(\lambda_N)$
and $z_2=   \lambda_1 -  \frac{1}{N}   b(\lambda_1)$.
The equations for the detached eigenvalues $\lambda_1$ and $\lambda_N$ read
\beq
 V'(\lambda_N) =   2 S_-(\lambda_N) + \frac{1}{a(\lambda_N)}
\eeq
\beq
 V'(\lambda_1) =   2 S_+(\lambda_1) - \frac{1}{b(\lambda_1)}
\eeq
and the equation for $z_1$ and $z_2$ become:
\beq
\beta_1 = S_-(\lambda_N) + \frac{1}{a(\lambda_N)} = S_+(\lambda_N)
\eeq
\beq
\beta_2 = S_+(\lambda_1) - \frac{1}{b(\lambda_1)} = S_-(\lambda_1)
\eeq
therefore they are both read from the analytic continuation at low temperatures
of the inverse of the Stieltjes and the solution (\ref{Fo}) holds at all temperatures.

 \subsection{Low temperature of the annealed model and large deviations}

In this section we make a digression to emphasize that two models may have essentially the same quenched
properties, but very different low-temperature annealed behavior. This may come as a surprise, given that the latter
is just the analytic continuation of the high-temperature phase: the continuations of two almost identical high temperature
behaviors become widely different at sufficiently low temperatures.

Consider a Gaussian $V_G(x)=x^2/2$ matrix model, and a slightly perturbed one $V_{NG}(x)=x^2/2+g/4 x^4$. 
The Stieltjes transforms are $S_{G}(x)=\frac{1}{2} (x-\sqrt{x^2-4})$ and
 $S_{NG}(x) = \frac{1}{2} (x+g x^3 - (1+g x^2 + 2 g a) \sqrt{x^2-4a^2})$ respectively \cite{brezin1993planar}
(here $a$ is the solution of $3 g a^4+a^2-1=0$).
For small $g$, $S_G$ and $S_{NG}$ do not differ very much at any $x$. The density of eigenvalues is also almost the same. 
If we now move to the  low temperature annealed problem, one eigenvalue detaches, 
and  enters the region of large $x$,
where the behavior  $V_{G}' \propto  x$ for the Gaussian and   $V_{NG}' \propto x^3$ for the perturbed one play an important role,
determining the behavior of $S_{G+}$ and $S_{NG+}$, the expressions of $S_{G}$ and $S_{NG}$ with a positive sign in front of the squared root,
considered for $x>>1$.
This means that the two models while behave very similarly in the high temperature phase (their energy scales as $E_{G}^{T\gg1} \sim E_{NG}^{T\gg1} \sim-\beta=-1/T$), 
they are very different
at low temperature, where the Gaussian model has an energy that goes as $E_G^{T\ll 1}(T) \sim -1/T$ while for the quartic potential
$E_{NG}^{T\ll 1}(T)\sim - T^{-1/3}$. As we shall see below, this effect manifests in the probability tails of a diagonal matrix
element drawn from a distribution with a potential $V_{NG}$, which scales as: $\ln P(A_{ii}) \propto -N A_{ii}^4$,
for large $A_{ii}$.

\subsection{Complex annealed landscape: the case $p>2$}\label{Sec_psin}

Let us now  generalize the previous results to a problem with a more complex landscape. 
The simplest of them all is a spherical model with  $p$-body interactions $p>2$:
\beq\label{Annealed_ising_pspin00}
E_J^p[ \mathbf{s}] = - \sum_{i_1...i_p} J_{i_1...i_p} s_{i_1}...s_{i_p} \;   \qquad ; \qquad 
\sum_i s_i^2= N
\eeq 

For  Gaussian couplings the model is well-studied. The ensemble of $J_{i_1...i_p}$ is rotationally invariant in this case, it is a tensor generalization
of a Gaussian matrix model. Non-Gaussian variants are also possible, although we shall not study them here.
From now on we concentrate on the case $p=3$.

The quenched solution has a phase transition 
while the annealed one 
\beq
\langle Z_J^p(\beta) \rangle = \int {\rm d} J P(J) \int  {\rm d} \mathbf{s} \
e^{- \beta E_J^p[ \mathbf{s}] }\;  \delta\left(\sum_i s_i^2 - N\right) \ .
\eeq
(superficially) does not \cite{Crisantisommers}.

The notion of  `detaching an eigenvalue' may be expected to generalize  as the fact that below the (quenched)  critical temperature, the annealed model  develops a `spike' in the interaction $J_{ijk} \rightarrow J'_{ijk} + \frac{a(T)}{N^2} \; v_i v_j v_k$, where the $J'$ have the same statistical properties as the $J$ 
(and are of $O(N^{-1})$), and  $a(T)$ and $v_i$ are of order one.
We have evidence for this of two sorts: dynamical and static. 

{\em i)} The dynamic case has actually been done for us
many years ago \cite{barrat1996dynamics}: Barrat {\it et al.} computed the evolution starting from an equilibrated configuration, obtained from
an annealed ansatz, as befits a system above the static transition temperature,  the Kauzmann (replica symmetry breaking) temperature $T_k$.
 In the high  temperature phase, the dynamics were completely ergodic, while in a regime of intermediate temperatures $T_k<T<T_d$ between the static and the dynamic transitions, it was found that, as expected,
the system is confined  to one of the many states that together constitute the Gibbs measure in that regime. 
What is relevant for us here is that  if we extend their annealed  solution below
 the static transition of the quenched system, annealed and quenched  measures
start do differ.  The initial annealed state is now a condition that corresponds to biased $J$'s, and the subsequent dynamics allows us to 
study its characteristics.  The dynamics is now inside a deep state whose energy and free-energy
have been `pulled below' the level of the  ground state for typical $J$'s. The `size' of this {\em self-planted}  state, as measured by the Edwards-Anderson parameter  of the dynamics, is small. All characteristics
of the state are  analytic continuations of the equilibrium states contributing to the Gibbs measure in the intermediate temperature phase.

{\em ii)} Once we know that the state generated by the annealed measure is deep, we still do not know whether it is isolated in phase-space, or a whole 
cluster of nearby states are `planted' by the annealed measure. 
To answer this, there is a static argument, which we shall develop in Section \ref{Sec_witness}: it is to show that once we have an annealed $J$ with a deep self-planted ground state, the landscape restricted to configurations orthogonal to that state is the same as the one of the old quenched system. In this sense,  the `spike' is isolated.

 \subsection{The  case of disorder on a lattice}
 
 Here we consider a generalization of annealed averages to finite dimensional models.
 Here, in general, the condition of rotational invariance
of the coupling matrix   ${\cal{P}}(J) = {\cal{P}}(U J U^\dag)$ for an orthogonal matrix $U$  is not satisfied.
When the interactions have a lattice structure that is not destroyed by annealing, there are clearly limitations on the spectra
as the ferromagnetic lattice can not be realized with a rank one perturbation.
The case of a three-dimensional Edwards-Anderson model $E=-\sum_{n.n.} J_{ij}s_i s_j$ with $J_{ij}=\pm 1$ 
and $s_i$ Ising spins is very clear. 
All ground states are configurations of the form  $J_{ij} =\sigma_i \sigma_j$, for a given set $\sigma_i = \pm 1$.   
This is a Mattis system, a ferromagnet in disguise, as one may check by gauge-transforming the spins of the system as $s_i \rightarrow \sigma_i s_i$.

 \subsection{Diffusion over annealed solutions}
 
Slow dynamics can arise when the system  has two different sets of
spins. In this Section we study numerically the problem of Ising spins, instead of the spherical ones, showing that
similar results hold also in this case.
The optimal solution may need to plant one or two deep valleys: if the optimal number is only one, the solution with two
valleys is metastable, but may be long-lived.
The partition function that we  analyze is then the following:
\beq\label{Annealed_ising_pspin}
\langle Z^p_J(\beta_1,\beta_2) \rangle = \int {\rm d} J P(J) \sum_{\boldsymbol{\sigma}^1,\boldsymbol{\sigma}^2 \in \{-1,1\}^N} 
e^{ \beta_1 \sum_{i_1...i_p} J_{i_1...i_p} \sigma_{i_1}^1..\sigma_{i_p}^1  + \beta_2 \sum_{i_1...i_p} J_{i_1...i_p} \sigma_{i_1}^2...\sigma_{i_p}^2} \ .
\eeq
and for simplicity we take $J_{i_1...i_p}$ Gaussian distributed with average zero and variance $\langle J_{i_1...i_p}^2\rangle= p!/2 N^{p-1}$.
The case $p=2$ corresponds to the replicated annealed SK model.
A direct annealed  computation of (\ref{Annealed_ising_pspin}) gives 
the following expression: 
\beq\label{Replica_pspin}
\begin{array}{l}
\displaystyle
\langle Z^p_J(\beta_1,\beta_2) \rangle = \int{\rm d} q \ 
e^{N \left[ \frac14 (\beta_1^2  + \beta_2^2 +  2 \beta_1 \beta_2 q^p) -\frac{1+q}{2} \ln \left(\frac{1+q}{2}\right)-\frac{1-q}{2} \ln \left(\frac{1-q}{2}\right)\right] } 
\ ,
\end{array}
\eeq
where $q = \sum_i \sigma_i^1 \sigma_i^2$ is the overlap between replicas.

The solution of this problem for $p=2$  predicts a second order phase transition
at $\beta_1\beta_2 = 1$ from $q=0$ to $q\neq0$. 
Note however that this solution does not say anything about the spectrum of the 
matrix $J$ in these two solutions.

\begin{figure}
\centering \includegraphics[angle=0,width=6cm]{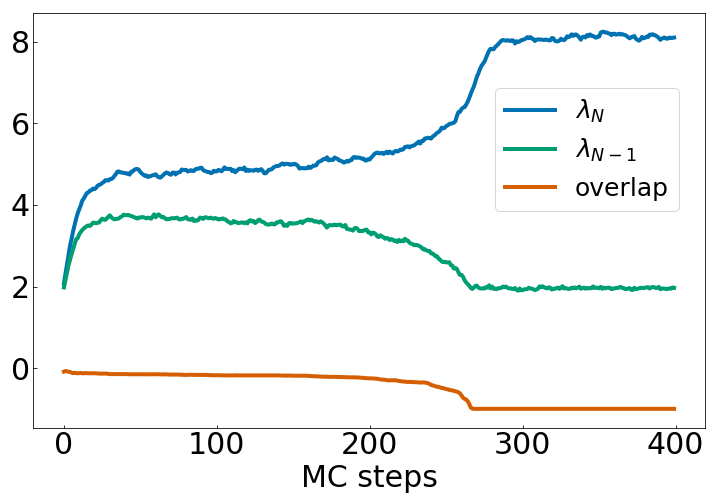}
\caption{A run with two replicas at temperature $\beta_1=\beta_2=4$ for $p=2$.
We show the first two eigenvalues and the overlap between the two replicas.}
\label{Fig_run_ising}
 \end{figure}

\begin{figure}
\centering \includegraphics[angle=0,width=6cm]{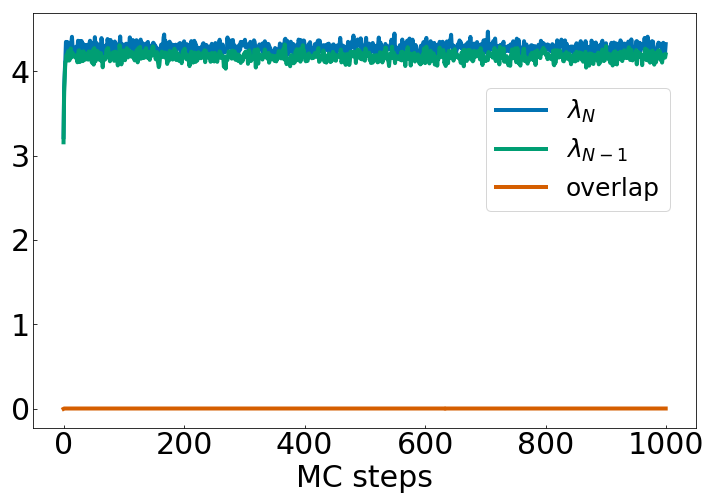}
\caption{A run with two orthogonal replicas at temperature $\beta_1=\beta_2=4$ for $p=2$.
We show the first two eigenvalues and the overlap between the two replicas.}
\label{Fig_run_orthogonal}
 \end{figure}

In Figure \ref{Fig_run_ising} we show a run of a Montecarlo simulation
of the model (\ref{Annealed_ising_pspin}) for $p=2$ with $N=500$ spins 
at low temperature ($\beta_1=\beta_2=4$), 
starting from two uncorrelated configurations.
In particular we present the first two eigenvalues: clearly at the beginning
of the the simulation two eigenvalues detach form the bulk $\lambda \in [-2,2]$ signalling
a metastable solution with low overlap between the configurations. 
However after a certain number of sweeps
the true equilibrium solution is reached with only one eigenvalue out of the bulk
and a perfect overlap between the two configurations 
 (which can be positive or negative with equal probability).
  We note that the solution with two eigenvalues out of the bulk is the stable one if
  one forces the two sets of spins to be orthogonal.
 In Fig. \ref{Fig_run_orthogonal} we show a run in the same condition as before but with 
 an additional coupling $\alpha \left( \sum_{i} \sigma_i^1 \sigma_i^2 \right)^2$ with $\alpha>>1$
 which does this.
We see that the solution with two eigenvalues out of the bulk
 and (obviously) small overlap is the equilibrium stable one in this case.

\begin{figure}
\centering \includegraphics[angle=0,width=6cm]{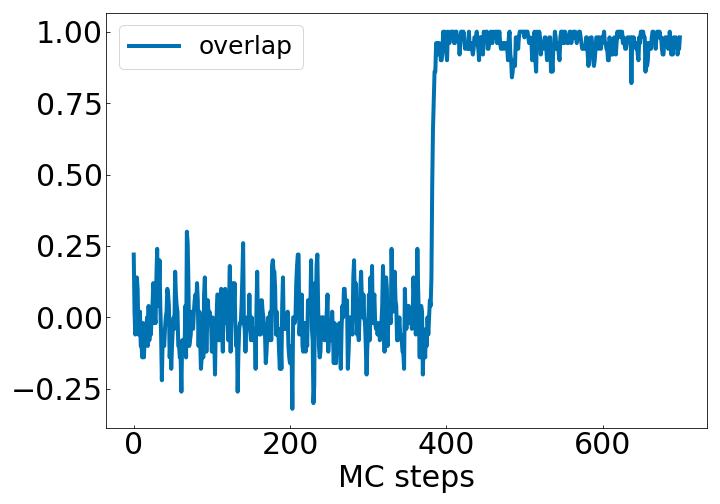}
\caption{A run with two replicas in the $p$-spin model with $p=3$ at temperature $\beta_1=\beta_2=1.2$.
The overlap between the two replicas stays close to zero until it jumps suddenly to one.}
\label{Fig_run_pspin}
 \end{figure}

For $p>2$ the situation is different. At low temperatures there are two phases separated by a first order phase transition depending on $|\beta_1\beta_2|$,  separating a phase with $q=0$
where there are no planted states or two orthogonal ones, and one with $q$ near one where both replicas are in the same planted state.
In Fig. \ref{Fig_run_pspin} we monitor the annealed dynamics of the overlap  (for $N=100$ spins) 
at  $\beta_1=\beta_2=1.2$ and we see that  it stays 
close to zero and then suddenly jumps to one, when both copies have `found one another'.

%
%
%

\subsection{`Witness' model and order parameter}\label{Sec_witness}

The free energy (\ref{Eq_p2_2}) is analytic for all $\beta$, and the same is true for all $p$. One expects, however,
that some discontinuity shows up at the point in which an eigenvalue suddenly detaches from the distribution, or that
a self-planted state appears in the more general situation. 
When one considers what is happening to the  distribution of the $J$'s, the strategy becomes obvious: generate the $J$'s with the annealed
process, and use them as the {\em quenched} disorder of a ``witness" model: either by studying its equilibrium, its dynamical properties, or
by  performing a Kac-Rice study of the saddle points of its potential.
An instance of this was already proposed in Ref. \cite{matsuda1984hidden}. Consider the distribution $\mathcal{P}(J)$ of $J$'s derived from an annealed process
(\ref{Eq_p2}): 
\beq
\mathcal{P}(J) \propto  P(J) \int {\rm d} \mathbf{s} \ e^{\frac12 \beta  \sum_{ij}J_{ij} s_i s_j } \delta\left(\sum_i s_i^2 - N\right)/\bar Z
\eeq where $\bar Z$ is the normalization,
and use it as quenched disorder for a system with spins $\sigma_i$:
\beq\label{Eq_p22}
Z(\beta) =  \frac{1}{\bar Z} \int {\rm d} J \ P(J)  \int {\rm d} \mathbf{s} \ e^{\frac12 \beta  \sum_{ij}J_{ij} s_i s_j } \delta\left(\sum_i s_i^2 - N\right) \ln \left\{
 \int {\rm d} \boldsymbol{\sigma} \ e^{\frac12 \beta  \sum_{ij}J_{ij} \sigma_i \sigma_j }  \delta\left(\sum_i \sigma_i^2 - N\right) \right\}
\eeq
It is clear that a detached eigenvalue in the $J$'s acts as a ferromagnetic (or rather, a Mattis term $(\sum_i v_i \sigma_i)^2$ ) for the $\sigma_i$, and this will show up in the inter-state correlation. In replica language, Eq (\ref{Eq_p22}) may be expressed as:
\beq\label{Free_energy_non_anal}
Z(\beta) \propto \langle Z_J(\beta) \log Z_{J}(\beta) \rangle =\left. \Big\langle\frac{\partial}{\partial n} Z_J^n \Big\rangle\right|_{n=1}
\eeq
and the order parameter:
\beq
q_{o} = \langle s_a s_b \rangle \;\;\;   a \neq b
\eeq
where the problem has $n \rightarrow 1$ replicas.
Exactly the same strategy may be used for complex landscapes, for example $p>2$, where there will be a transition to a deep  planted state \cite{zdeborova2016statistical}.
The strategy  may be implemented with two or more families of $\sigma_i$, if more self-planted states are to be detected.

Let us now make the calculation we announced in Section \ref{Sec_psin} for $p=3$ to show that there are no deep states
orthogonal to the planted state. We demand  that the spins in the `witness' model be orthogonal to those
in the planted one, and check that with such restriction the model is the same as the quenched one:
\beq\label{Annealed_ising_pspin0}
\begin{array}{l}
\displaystyle
\tilde{Z}(\beta) = \int {\rm d} J P(J)   \int {\rm d} \mathbf{s} \
e^{\beta \sum_{i<j<k} J_{ijk} s_i s_j s_k }\;  \delta\left(\sum_l {s}_l^2 - N\right) 
\\ \vspace{-0.2cm} \\
\displaystyle
\qquad\qquad
\ln \left[ \int {\rm d} \boldsymbol{\sigma} \
e^{ \beta \sum_{i<j<k} J_{ijk} \sigma_i \sigma_j \sigma_k }\;  \delta\left(\sum_l {\sigma}_l^2 - N\right) \delta\left(\sum_l \sigma_l s_l \right) \right] \ .
\end{array}
\eeq
Again, this is problem with $n+1$ replicas, with $n \rightarrow 0$. The constraint that imposes that the $\sigma_i$ be orthogonal
to the $s_i$ means, once the replica trick is applied, that the matrix $Q_{ab}$ will have $Q_{1a}=Q_{a1}=0$ so that the replica matrix breaks into a one by one and
an $n\times n$ block.  It is  a general property of the replica trick that when the matrix breaks into blocks the replicas uncouple.
Hence we get the original  annealed problem for $Q_{11}$ and an ordinary quenched one for $Q_{ab}$ ($a>1, b>1$). 
Thus, we conclude that the model restricted
to be orthogonal to the $s_i$ vector is just as if the $J$ were drawn from quenched ensemble.

\section{The distribution of elements in matrix models}\label{Sec_matrix}

We wish to compute the joint distribution of a number of elements $\{A_{a_1b_1}...A_{a_sb_s}\}$  of a matrix model
with invariances ${\cal{P}}(A) = {\cal{P}}(U A U^\dag)$, and we consider here the case where
$A$ is real and symmetric and $U$ is an orthogonal matrix. 
The generalization to the complex case is also possible.
The most usual case is when  the  model is defined through: ${\cal{P}}(A)= e^{-N {\mbox{tr}} W(A)}$
 (playing  the role of the distribution  ${\cal{P}}(J)= e^{- \frac{N}{2} {\mbox{tr}} V(J)}$ in the previous section).
 Let  $\hat A_{ab}$ be an  $r\times r$ submatrix, where $a$ and $b$ may be assumed to be $1,...,r$.
In the following we will denote with a symbol $\,\hat{}\,$ all matrices of small $r=O(1)$ size. 
  In \cite{foini2019eigenstate} we showed that the probability distribution of a sub-block
 of size $2\times 2$ 
 inherits the property of rotational invariance of the original ensemble, namely
 that 
 \beq\label{Prop_rot_inv}
P(\hat A) = P(\hat U \hat A \hat U^\dag)
 \eeq
where now $\hat U$ are $2\times 2$ elements of the group.
Below we will argue that the same proof can be generalized to a block of arbitrary dimension, 
and we shall relate this calculation of $P(\hat A)$ 
to the problem of solving annealed spin-glass models.

 \subsection{Probability distributions and spin-glass models}
 
 Let us start, for clarity, with the case $r=1$.
We consider the probability distribution:
\beq\label{Eq_P_r1}
 P_{r=1}(\hat A_{11})  = \int \; {\rm d} A \;  {\cal{P}}(A) \;  \delta(\boldsymbol{\sigma} \cdot A \boldsymbol{\sigma}  - \hat A_{11}) 
\eeq 
where $\boldsymbol{\sigma}$ is a $N$-component vector with unit norm. 
Eq. (\ref{Eq_P_r1}) corresponds to the diagonal matrix element of the matrix $A$ in a basis where $\boldsymbol{\sigma}$ 
in one of the vectors defining that basis. As the ensemble is invariant under change of basis such quantity does not
depend on $\boldsymbol{\sigma}$ and we can average over that:
 \begin{equation}\label{Eq_A11}
 P_{r=1}(\hat A_{11}) = {\cal{N}}  \int \; {\rm d} A \;  {\rm d} \boldsymbol{\sigma} \;  {\cal{P}}(A) \;  \delta(\boldsymbol{\sigma} \cdot A \boldsymbol{\sigma}  - N  \hat A_{11}) \delta(\boldsymbol{\sigma} \cdot \boldsymbol{\sigma} - N)
 \end{equation}
 where the constant ${\cal{N}} $ ensures the normalization of the probability, and we have rescaled the vector $\boldsymbol{\sigma}$ to have norm $N$ for convenience.
 The problem of the calculation of one diagonal matrix element has become the calculation of the annealed entropy of a spin-glass, its Laplace transform will be the partition function we studied in the previous section.

Let us now generalize to the probability distribution of a sub matrix of size $r\times r$. 
We have:
\begin{eqnarray}
 P_r(\hat A) &=& {\cal{N}}   \int \; {\rm d} A \; \prod_a {\rm d}   \boldsymbol{\sigma}_a \;  {\cal{ P}}(A) \; 
 \delta(\boldsymbol{\sigma}_a \cdot A \boldsymbol{\sigma}_b  - N \hat A_{ab}) \prod_{a b} \delta(\boldsymbol{\sigma}_a \cdot \boldsymbol{\sigma}_b -N \delta_{ab})\nonumber\\
  \label{micro}
 \end{eqnarray}
 where, again, we used the invariance under change of basis to average over the $\boldsymbol{\sigma}_a$.
 This is precisely a spin-glass model with $r$ sets of spins, forced to remain orthogonal.
 The fact mentioned above that  $P(\hat A)$ is invariant with respect to transformations in the $r \times r$ space implies that $P(\hat A) $ is a function
 of the eigenvalues of $\hat A$.  This invariance can be verified directly in Eq (\ref{micro}) by applying a similar reasoning as that
 described in \cite{foini2019eigenstate}.
  In particular one can consider $\hat U$ in (\ref{micro}), where $\hat U$ acts non trivially on the basis $\boldsymbol{\sigma}_a$. 
  Upon change of variables, and identifying $\hat U$ with a $N\times N$ orthogonal matrix that acts as $\hat U$ on the space defined by  $\boldsymbol{\sigma}_a$
  and leaves unperturbed the other $N-r$ vectors of the basis,  one indeed checks that $P_r(\hat U \hat A \hat U^\dag)=P_r(\hat A)$.

  One can also write, for ${\cal{ P}}(A) \sim e^{-N \text{tr} W(A)}$: 
 \begin{equation} \label{formula1}
P_r(\hat A) = {\cal{N}}   \int \; {\rm d} A \; e^{-N \text{tr} W(A)} \;  \prod_a {\rm d}   \boldsymbol{\sigma}_a \; 
\prod_{a b} d\beta_{ab}  dz_{ab} \; e^{\frac12 \beta_{ab}(\boldsymbol{\sigma}_a \cdot A \boldsymbol{\sigma}_b  - N \hat A_{ab} )- \frac12 z_{ab} (\boldsymbol{\sigma}_a \cdot\boldsymbol{\sigma}_b -N \delta_{ab})}     
\end{equation} 
where the integrals over the $\beta_{ab}$ and $z_{ab}$ run over the imaginary axis. 
 
 We may now explicitly diagonalize $\beta_{ab}=\sum_c \hat U^{\beta}_{ac} \tilde \beta_c \hat U^\beta_{bc}$, $z_{ab}=\sum_c \hat U^{z}_{ac} \tilde z_c \hat U^z_{bc}$ and
 $\hat A_{ab}=\sum_c \hat U^{A}_{ac}\tilde a_c \hat U^A_{bc}$ and integrate first over the parameters defining $\hat U^\beta$ and $\hat U^z$. The saddle
 point over those happens when $z_{ab}$ and $\beta_{ab}$ are both diagonal in the same basis as  $\hat A_{ab}$.
Integrating the $\boldsymbol{\sigma}_a$ away, in terms of the eigenvalues $\lambda_i$ of $A$, the exponent in Eq (\ref{formula1}) becomes,
distinguishing the $r$ eigenvalues $\tilde \lambda_1,...,\tilde \lambda_r$ that may be detached from the bulk:
 \begin{eqnarray}
 &\,&  - \, \,N \sum_{i \in {\mbox{\tiny bulk}}} W(\lambda_i) -N \sum_{k=1}^r W( \tilde \lambda_k) + \sum_{ij \in {\mbox{\tiny bulk}}} \ln |\lambda_i-\lambda_j| + \sum_{k,k'=1}^r \ln |\tilde \lambda_k-\tilde \lambda_k'|  \qquad \qquad \qquad \nonumber \\
 \qquad &\qquad +&  \sum_{k=1}^r \sum_{i \in {\mbox{\tiny bulk}}} \ln |\lambda_i-\tilde \lambda_k| -\frac N2  \sum_{k=1}^r \sum_{i \in {\mbox{\tiny bulk}}} \ln(\tilde z_k-\lambda_i \tilde \beta_k) + \frac{N}{2}\sum_{k=1}^r \tilde z_{k}  \nonumber \\
\qquad  &\qquad -& \frac{N}{2} 
 \sum_{k,k'=1}^r \; \ln (\tilde z_k  -  {\tilde \beta_k \tilde \lambda_{k'}} ) - \frac{N}{2} \sum_{n=1}^r \tilde \beta_k \tilde a_k
\end{eqnarray}
We recognize three kinds of terms:
\begin{itemize} 
\item a bulk term, just as if the system were quenched, but with (at least) $N-r$ eigenvalues
\item a subdominant term of interaction between detached eigenvalues of $O(1)$
\item a sum of terms for each detached eigenvalue (and its corresponding $\tilde z_n$) with the same form
of an $r=1$ problem formula.
\end{itemize}
The first contribution gives an $O(1/N)$ correction to the density of eigenvalues of the bulk. The interaction between detached eigenvalues
only acts if their differences are $O(1/N)$: it is a correction that couples real replicas. The third contribution is what we are concerned with: it is {\em superficially} 
(because these terms potentially interact through their effect on the bulk of $\lambda_i$) 
a sum of noninteracting terms for each replica, each of the same form of the one of $r=1$. 
If we compute the saddle point for each $\tilde z_a$, and eliminate then to obtain a form, in terms of the probability 
of a diagonal term $P^{\text{diag}}(\hat A_{11})=P_{r=1}(\hat A_{11})$:
\begin{equation}
\ln P_r( \hat A) = \sum_{a=1}^r  \ln P^{\text{diag}}(\tilde a_n) = - \frac{N}{2} \;  \text{tr} \, \tilde W( \hat A )
\label{final}
\end{equation}
where we have defined $\tilde W(x) \equiv - \frac{2}{N} \ln P^{\text{diag}}(x)$.  In short: {\em the large deviation function of an $r\times r$  submatrix}  {tr} $\tilde W(\hat A)$ {\em  is the same for all finite $r$}.

\subsection{Generating functions and marginals}

Let us write the Laplace transform of Eq. (\ref{formula1}) as
\begin{eqnarray}\label{Generating_funct}
 \langle e^{\frac{N}{2} \text{tr} \hat \beta \hat A}\rangle  \sim \int  \prod_a {\rm d}   \boldsymbol{\sigma}_a  \; dz_{ab} \; {\rm d} A \; {\cal{P}}(A) \; \prod_{ab}  e^{\frac12 \hat\beta_{ab}\boldsymbol{\sigma}_a \cdot A \boldsymbol{\sigma}_b -  \frac12 z_{ab} (\boldsymbol{\sigma}_a \cdot  \boldsymbol{\sigma}_b  - N \delta_{ab})}
 = \langle Z_A(\hat \beta) \rangle \sim  e^{\frac{N}{2} \Phi(\hat \beta)} 
\label{unaveraged}  
\end{eqnarray}
where the first brackets means $\langle \bullet \rangle = \int D \hat A \bullet P_r (\hat A)$.
Eq. (\ref{Generating_funct}) is the generating function of Eq. (\ref{formula1}).
The passage of (\ref{micro}) to (\ref{unaveraged}) is the standard one between microcanonical to canonical, except
that there are several `temperatures'.
We recognize the {\em annealed average over ``disorder" (=$A$)} of  the spherical model with $r$  `replicas' {\em that are forced to be orthogonal},
the quantity $\langle Z_A(\hat \beta) \rangle$.
The thermodynamic relations are (for $N \rightarrow \infty$):
\begin{eqnarray}
\hat A_{ab}&=&   \frac{\partial \Phi}{\partial \beta_{ab}} \qquad \qquad  \qquad \qquad
\beta_{ab} = \frac{\partial \ \text{tr} \, \tilde W( \hat A ) }{\partial \hat A_{ab}} \label{eqqq}
\end{eqnarray}
The quantity $e_{ab} = - \hat A_{ab}$ plays the role of energies, the $\beta_{ab}$ of temperatures, and $s(\hat A) = \text{Const} - \text{tr} \, \tilde W(\hat A)$ of entropy.
Eliminating the temperatures
 $\beta_{ab}$ in favor of the $\hat A_{ab}$ one obtains the desired distribution  $ \ln P_{r}(\hat A)$.  
 This is particularly clear if we think to the case $r=1$.
 
\begin{figure}
\begin{center}
\includegraphics[width=5.8cm]{Figs/energy11.pdf} \hspace{1cm} \includegraphics[width=6.4cm]{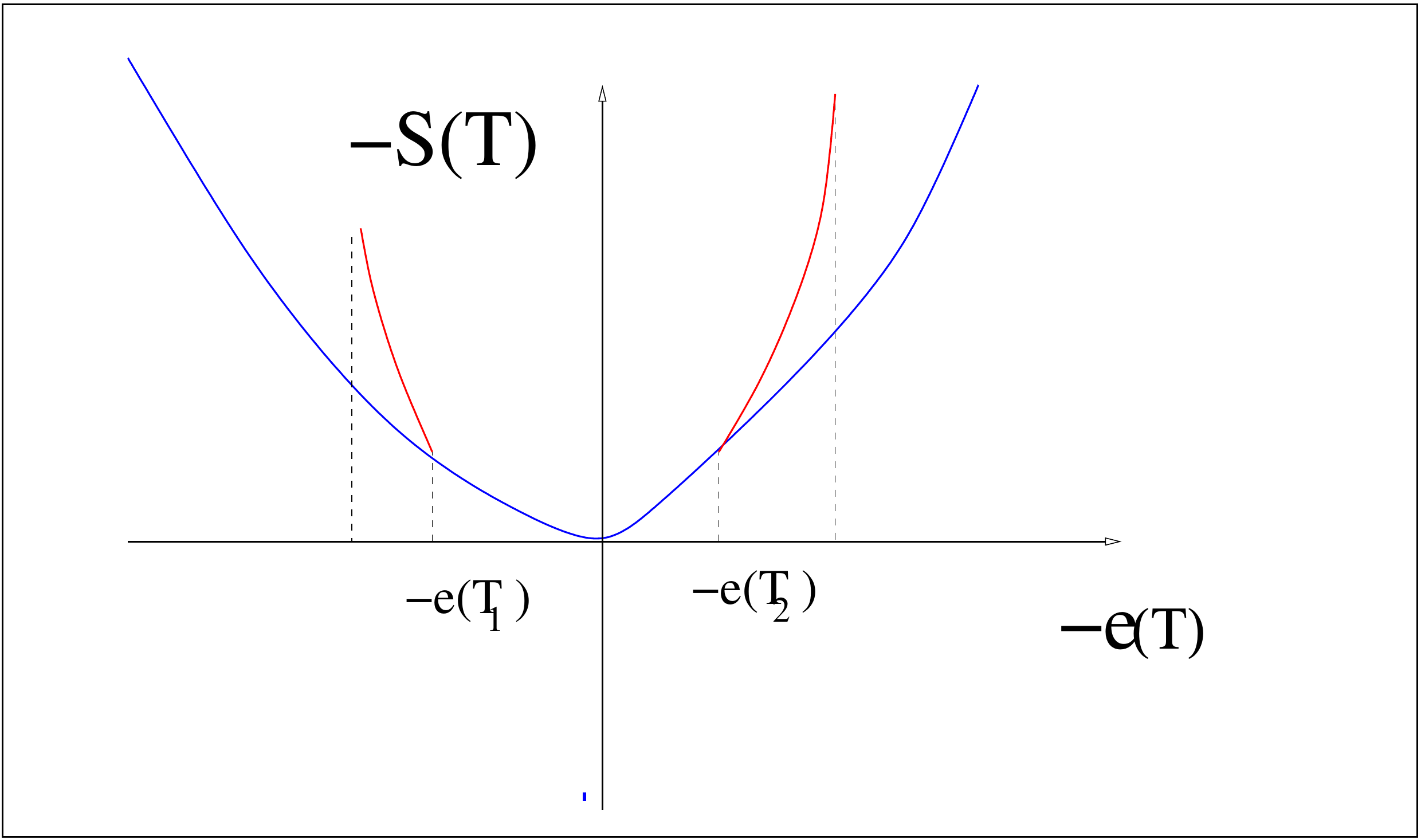}
\caption{Left: a sketch of an energy versus temperature ($T=\beta^{-1}$) curve for the annealed and quenched problem. 
Right: the corresponding parametric plot of $-s(T)=- \frac{\partial T \Phi }{\partial T}$ 
versus $-e(T)$ which gives the large deviation function $\tilde{W}(\hat A)$ for the probability $P \sim e^{- \frac{N}{2} \tilde{W}(\hat A)}$ of a diagonal matrix element $\hat A=-e$. The range of the quenched probability is bounded, while the 
one of the annealed problem is not.}\label{Fig_energy_T}
\end{center}
\end{figure}

The relation between thermodynamics and the probability of diagonal matrix elements is summarized in Figure
\ref{Fig_energy_T}.

If we wish to calculate the marginal distribution of some elements of the $r \times r$ matrix, we need to integrate (\ref{formula1}) over some of the $\hat A_{ab}$.   This, in turn, amounts to setting to zero the corresponding $\beta_{ab}$ in Eq. (\ref{unaveraged}). The remaining variables,
those not integrated upon, are given by Eq (\ref{eqqq}).
For example, if we wish to compute the joint distribution of $r$ diagonal elements $\hat A_{aa}$, with $a=1,...,r$, we set $\beta_{ab}=0$ for $a\neq b$. 
Another example is to calculate the probability distribution of an off-diagonal element, we set $r=2$ and $\beta_{11}=\beta_{22}=0$, $\beta_{12}=\beta_{21}=\beta$.
We shall do this below.

\subsection{Diagonal matrix elements in matrix models}\label{Sec_diagonal}

The marginal joint distribution of $r$ diagonal eigenvalues is obtained by setting $\beta_{ab} =0$ for $a\neq b$.
The analysis  is just as above.
We thus conclude that, to leading order, the marginal:
\begin{equation}
\ln  P(\hat A_{11}, ... , \hat A_{rr}) = \sum_{a=1}^r \ln P^{\text{diag}}(\hat A_{aa})
\end{equation}
i.e. the diagonal elements are independent for finite $r$. Let us emphasize  that this is a large $N$ result, valid for $r$ much smaller than $N$,
such that the repulsion between eigenvalues may be neglected.

\subsection{Off-diagonal matrix elements: two replicas at opposite temperature}\label{Sec_off}

Considering $r=2$ with equal off-diagonal ``temperatures" $\beta_{12}=\beta_{21}=\beta$ and
zero diagonal terms  $\hat \beta_{aa}=0$ in (\ref{unaveraged}) as indicated above, we get:
\begin{eqnarray}
\langle Z_A^{\text{off}}(\beta)\rangle &\equiv&\int  \prod_a {\rm d}   \boldsymbol{\sigma}_a  \; \prod_{a b} dz_{ab} \;  {\rm d} A \; {\cal{P}}(A) \;   e^{\frac12 \beta \boldsymbol{\sigma}_1 \cdot A \boldsymbol{\sigma}_2 + \frac12 \beta \boldsymbol{\sigma}_2 \cdot A  \boldsymbol{\sigma}_1}\; \prod_{ab} e^{- \frac12 z_{ab} (\boldsymbol{\sigma}_a \cdot  \boldsymbol{\sigma}_b  - N \delta_{ab})}
\label{unaveraged3}  
\end{eqnarray}
The symbol $\langle Z_A^{\text{off}}(\beta)\rangle$ is indicating, as in (\ref{Generating_funct}), 
that such generating function for off-diagonal matrix elements,
is equivalent to an annealed average of an associated spin glass problem.
In fact, a rotation to $(\boldsymbol{\sigma}_1\pm \boldsymbol{\sigma}_2)/\sqrt 2$, and similarly for the $z_{ab}$ leads to:
\begin{eqnarray}
\langle Z_A^{\text{off}}(\beta)\rangle &\equiv&\int  \prod_a {\rm d}   \boldsymbol{\sigma}_a  \; \prod_{a b} dz_{ab} \; {\rm d} A \; {\cal{P}}(A) \;   e^{\frac12 \beta \boldsymbol{\sigma}_1 \cdot A  \boldsymbol{\sigma}_1 - \frac12 \beta \boldsymbol{\sigma}_2 \cdot A  \boldsymbol{\sigma}_2}\; \prod_{ab} e^{- \frac12  z_{ab}(\boldsymbol{\sigma}_a \cdot  \boldsymbol{\sigma}_b  - N \delta_{ab})}
\end{eqnarray}
and therefore one sees that the problem of the generating function of an off-diagonal matrix element 
maps into two orthogonal  replicas with opposite temperatures.
 Performing the same steps as in the previous section one arrives
to the partition function:
\beq
\langle Z_A^{\text{off}}(\beta)\rangle \propto \int {\rm d} z_1 {\rm d} z_2 {\rm d} z_{12} e^{\frac{N}{2} ( z_1+ z_2 + \sum_k
\log \left[ (z_1- \beta\lambda_k)(z_2+\beta\lambda_k) - z_{12}^2 \right]}
\eeq
The saddle point over $z_{12}$ which enforces the orthogonality between the two vectors
 imposes $z_{12}=0$ and therefore the result is that of two independent
diagonal integrals, a particular case of what discussed in Section \ref{Two_replicas_two_T}
for replicas at opposite temperature (which leads at low temperature to two eigenvalues one to the left and one to the right 
out of the bulk).
The result (\ref{Fo}) implies the following relation between diagonal and off-diagonal matrix elements:
\beq\label{Relation_for_Z_off}
\langle Z_A^{\text{off}}(\beta)\rangle = \langle Z_A^{\text{diag}}(\beta)\rangle\langle Z_A^{\text{diag}}(-\beta)\rangle \ .
\eeq
where $\langle Z_A^{\text{diag}}(\beta)\rangle$ is the (annealed) partition function of one single replica at 
inverse temperature $\beta$, or equivalently $\langle e^{\frac12 \beta \hat A_{11}}\rangle$.

\subsubsection{Applications}

As an application let's see how these formulas allows us to compute the distribution
of matrix elements in some specific ensembles where we know the R-transform.
The result (\ref{High_T}) allows us to 
 write for the diagonal matrix elements:
\beq\label{Prob_diag}
P_{A_{ii}}(a) \sim e^{\frac{N}{2} \min_{\beta} \left[ - \beta a + \int_0^\beta {\rm d} x \ R(x)  \right] } 
\eeq
Form this, knowing that $R(x)=x$ of a Gaussian ensemble \cite{tulino2004random}, we get
\beq
P_{A_{ii}}^G(a) \sim e^{-N a^2/4}
\eeq
Moreover from (\ref{Fo}), (\ref{Relation_for_Z_off}) and:
\beq
P_{A_{ij}}(a) \sim e^{N \min_{\beta} \left[ - \beta a +  \frac12 \int_0^\beta {\rm d} x \ R(x) +  \frac12 \int_0^{-\beta} {\rm d} x \ R(x)  \right] } 
\eeq
we recover the  expected result for the off-diagonal matrix element of a Gaussian ensemble:
\beq
P_{A_{ij}}^G(a) \sim e^{- N a^2/2}
\eeq
Let us now analyze a less trivial case.
For a Wishart matrix $A=W W^\dag $ where $W$ is matrix of size $N\times K$ ($N \leq K$)
whose entries are Gaussian random variables with zero mean and variance $1/N$,
it is known that $R(x)=\alpha/(1-x)$  with $\alpha=K/N$ \cite{tulino2004random}.
With this we obtain the probability distribution of a diagonal matrix element of such matrix,
valid in the large $N$ limit:
\beq\label{Eq_Wishart_D}
P_{A_{ii}}^W(a) \sim e^{\frac{N}{2}  (- a + \alpha \log a)} \ ,
\eeq 
which could be generalized to the complex case as in \cite{zhang2017average}.
 For the off-diagonal matrix elements we derive the following probability distribution:
\beq
\begin{array}{lcl}
\log P_{A_{ij}}^W(a)  \displaystyle& \sim &
  \displaystyle - \frac{N}{2} \left[ \alpha  + \sqrt{4 a^2+\alpha^2} - \alpha \log\left(\frac{2a+\alpha + \sqrt{4 a^2 +\alpha^2}}{2}\right) \right.
\\ \vspace{-0.2cm} \\
&& \displaystyle \qquad\qquad\qquad\qquad\qquad \left.
- \alpha \log\left(\frac{- 2 a+\alpha + \sqrt{4 a^2 +\alpha^2}}{2}\right)
\right] \ .
\end{array}
\eeq

Let us finally note that the expression (\ref{High_T}) in terms of the R-transform, which can be 
written as a series expansion starting from (\ref{Eq_R_series}), establishes a relation between the cumulants 
of the random variable $A_{ii}$, the diagonal matrix element, or the cumulant of the off-diagonal matrix element $A_{ij}$
and the free cumulant $C_k$ of the matrix ensemble under consideration.

\section{Conclusions}

In this paper we study how an annealed measure `deforms' the disorder, or, equivalently, how large deviations depend on rare realizations
for the disorder. 
 These realizations mimic the `planted' ensemble, where the landscape is modified by creating an unusually deep valley,
   which is used in the theoretical study of inference problems,
but in the annealed setting this valley  is self-generated.
This study was initially motivated by the observation that  annealed computations are the most natural ones in spin-glass models that arise when one is interested in the
probability of matrix elements of a matrix generated with a  matrix model Hamiltonian. We discuss these in detail in the second part of the work,
where our results show that large deviations (the tails of the distribution)  of matrix elements occur when one or more 
eigenvalues\ have detached from the (typical) bulk of the spectrum of the matrix.

\section*{Acknowledgements}

We thank Davide Facoetti for useful discussions.  This work is supported by ``Investissements d'Avenir" LabEx PALM
(ANR-10-LABX-0039-PALM) (EquiDystant project, L. Foini). J.K. is supported by the Simons  Foundation Grant No 454943.

\appendix

\bibliography{Biblio}

\end{document}